\documentclass[aps,prd,reprint,groupaddress,amsmath,amssymb]{revtex4-2}
\usepackage{color}
\usepackage{amsmath,amssymb}
\usepackage{graphicx}
\usepackage{ulem}
\usepackage[colorlinks, linkcolor=blue, citecolor=blue, urlcolor=blue, breaklinks=true]{hyperref}
\usepackage{subfig}
\usepackage{mathtools}

\usepackage{mathpazo,times} 
\DeclarePairedDelimiter\bra{\langle}{\rvert}
\DeclarePairedDelimiter\ket{\lvert}{\rangle}
\DeclarePairedDelimiterX\braket[2]{\langle}{\rangle}{#1\,\delimsize\vert\,\mathopen{}#2}

\begin{document}
	\title{Quantum Transitions of Vector Vortex Light in  Gravitational Waves}
	\author{Haorong Wu}
	\affiliation{School of Physics and Technology, Wuhan University, Wuhan 430072, China}
	\author{Xilong Fan}
	\email{xilong.fan@whu.edu.cn}
	\affiliation{School of Physics and Technology, Wuhan University, Wuhan 430072, China}
	
	\begin{abstract}	
		We develop a theoretical framework to describe the full interaction between vector vortex light fields and gravitational waves (GWs). Using perturbation theory and the canonical quantization of the electromagnetic field, we calculate the quantum transition probabilities of vector Bessel beams propagating through  GWs. We demonstrate that GWs induce fourteen different quantum transition channels across orbital angular momentum (OAM) $l$ and spin angular momentum (SAM) $\sigma$, mapping initial states $\ket {\sigma,l}$ to $\ket {\sigma+\Delta \sigma,l+j-\Delta \sigma}$, where $\Delta\sigma \in \{-2, 0, 2\}$ represents the change in SAM and $j \in \{-3, \dots, 3\}$ denotes the change in total angular momentum. Among these channels, SAM-conserving transitions between OAM states, specifically $\ket {\sigma, l}\rightarrow \ket {\sigma, l\pm 1}$, provide the most viable mechanism for experimental detection. Conversely, spin-flip transitions are shown to be heavily suppressed relative to OAM transitions.	Additionally, the reversal of SAM induces an asymmetric shift in the OAM transition channels, reflecting the underlying coupling between SAM and OAM during the gravitational interaction. Based on these transition channels, we propose a new cavity-based GW detection configuration. By relying on quantum transitions rather than macroscopic arm-length changes, this scheme is inherently insensitive to displacement-based disturbances like seismic noise, offering a new paradigm and frequency bands for GW observation.
	\end{abstract}
	\maketitle
	
	\section{Introduction}
	The detection technique of current gravitational wave (GW) observatories (including LIGO, Virgo, and KAGRA) relies on Michelson interferometry \cite{Aasi_2015,Acernese_2015,PhysRevLett.123.231107}. In this scheme, passing spacetime perturbations alter the interferometer arm lengths, which breaks the destructive interference condition and translates the GW signature into optical intensity fluctuations at the output port. Information is thereby extracted through a phase difference between the light from the two arms. However, this interferometric techniques have not yet exploited the full parameter space of the light field, neglecting the angular momentum (AM) degrees of freedom. 
	
	Since the foundational work of Allen {\it et al.} \cite{allen1992orbital}, the spatial phase structure associated with photon orbital angular momentum (OAM) has driven advances across modern optics \cite{allen1992orbital,calvo2006quantum,mair2001entanglement,molina2001management,leach2002measuring,vaziri2002experimental,vaziri2003concentration,allen1999iv,kivshar2001optical,molina2007twisted,yao2011orbital,molina2001propagation}. This progress has recently motivated exploring the intersection of OAM and GW physics. Implementing higher-order Laguerre-Gauss modes in GW detectors, for instance, has been shown to substantially mitigate thermal noise \cite{PhysRevD.79.122002,PhysRevLett.105.231102}. Beyond noise reduction, the helicity coupling between structured light and GWs introduces novel physical effects. Wu {\it et al.} demonstrated that this coupling drives transitions between different photon OAM eigenstates, yielding newly measurable optical signatures \cite{wu2022testing}. The propagation of twisted photons through curved spacetime has also been shown to influence the coherence and degree of high-dimensional OAM entanglement within a GW \cite{PhysRevD.106.045023}. Conversely, structured light offers active control over GW generation in the high frequency regime, where laser pulse parameters and precisely designed optical arrangements can dictate the frequency, polarization states, and emission direction of the resulting GW \cite{PhysRevD.110.044023}.
	
	In our recent work, we investigated the evolution of scalar vortex light within  GWs and demonstrated that GWs induces measurable quantum transitions between quantized OAM states \cite{q38s-k2tq}. We calculated this interaction by applying the theory of wave propagation in linearized gravity alongside the canonical quantization of the scalar field in curved spacetime. Our analysis revealed that an incident photon with an initial OAM of $l$ interacting with a GW excites the $l\pm1$ and $l\pm2$ spatial modes with probabilities on the order of $P_{l\pm1}\sim 10^{-17}$ and $P_{l\pm2}\sim 10^{-20}$, respectively. Based on these transition probabilities, we previously proposed a new detection method designed to achieve high signal photon rates across a broad frequency spectrum. We showed that this design is insensitive to seismic noise and provides significant advantages over conventional interferometers for resolving the distance to astrophysical sources.
	
	Following our investigation into OAM, we also established how the spin angular momentum (SAM) of light behaves in a GW \cite{wu2026chir}. By extending our theoretical framework, we demonstrated that a GW actively flips photon chirality via AM exchange, an interaction we termed the spin-2-gravitation chiroptical effect. We derived the selection rules for this process, showing they are strictly dictated by the underlying spin-1 and spin-2 symmetries of the electromagnetic (EM) and gravitational fields, respectively. Because SAM is an inherently local property, it cannot accumulate gravitational perturbations over extended areas. Consequently, we found the magnitude of this chiroptical effect to be exceedingly small relative to corresponding OAM interactions with GWs. Despite this small magnitude, this effect can serve as a theoretical tool for probing the chiral structure of GWs.
	
	These previous findings naturally raise the question of how vector vortex light evolves in a GW and whether the inclusion of SAM modifies or amplifies the quantum transitions between OAM states. In this paper, we unify our prior theoretical frameworks to describe the full interaction between vector vortex light fields and GWs. This combination introduces difficulties arising from the gauge conditions required by the vector nature of the light field and the orthonormality property of basis states within a high-dimensional Hilbert space. To overcome these challenges, we will start from establishing the appropriate EM mode functions in Minkowski spacetime.
	
	Throughout this paper, we adopt the units with $c=G=\hbar=1$ and the metric signature $(-,+,+,+)$. Greek indices denote spacetime dimensions $\{0, 1, 2, 3\}$, while Latin indices represent spatial dimensions $\{1, 2, 3\}$. The function $\delta(x)$ represents the Dirac delta for continuous variables and the Kronecker delta for discrete ones. Cartesian and cylindrical coordinates are defined as $x^{\mu} = \{t,x,y,z \}$ and $x^{\tilde \mu} = \{t,\rho,\phi,z\}$, with corresponding coordinate basis vectors $\partial_{\mu}$ and $\partial_{\tilde \mu}$. To address the unnormalized nature of the cylindrical coordinate basis, we will also use the orthonormal noncoordinate basis $\{\hat t,\hat \rho,\hat \phi,\hat z \}$. Any index carrying a hat (for example, $A^{\hat \alpha}$) signifies an expansion within this cylindrical orthonormal basis.
	
	\section{Quantization of vector Bessel beams in Minkowski spacetime}
	Various approaches exist for constructing a quantum vector Bessel light field. For example, one can begin with the following ansatz for the mode functions \cite{PhysRevA.99.023845}
	\begin{equation}
		A^0_I(x^\alpha)=A^3_I(x^\alpha)=0,~~A^a_I (x^\alpha)=A^a_I(x,y)e^{i(-\omega t+k_3 z)},
	\end{equation}
	where the index $a$ represents transverse dimensions $\{1, 2\}$, $A^a_I (x^\alpha)$ satisfies the wave equation for EM fields, and $I$ represents a set of indices. For plane waves, $A^a_I (x,y)$ can be chosen to satisfy the Lorenz or Coulomb gauge. However, this same ansatz will not generally satisfy these gauges for a vortex beam. One can resolve this limitation by requiring that $A^\mu_I(x^\alpha)$ satisfy the gauge conditions while $A^a_I (x^\alpha)$ is retained. By selecting the radiation gauge, one derives
	\begin{equation}
		A^0_I(x^\alpha)=0,~~A^3_I(x^\alpha)_{,3}=-A^a_I (x^\alpha)_{,a}.
	\end{equation}
	For a Bessel beam, we define
	\begin{equation}
		A^a_I (x^\alpha)=CJ_m(k_\perp \rho)e^{im\phi}e^{i(-\omega t+k_3 z)}e^a_\sigma,
	\end{equation}
	where $C$ is the normalization factor, $J_m(\cdot)$ is the Bessel function of order $m$, $k_\perp$ is the radial wavevector, $m$ is the topological charge, $k_3$ is the wavevector along the $z$ axis, $\omega=(k_3^2+k_\perp^2)^{1/2}$, $\mathbf{e}_\sigma=(\mathbf{e}_x+i\sigma \mathbf{e}_y)/\sqrt 2$ is the polarization vector, and $I$ represents the index set $\{k_3,k_\perp,\sigma, m \}$. Then, $A^3_I(x^\alpha)$ is evaluated as
	\begin{equation}
		A^3_I(x^\alpha)=-i\sigma \frac{Ck_\perp}{\sqrt 2 k_3}J_{m+\sigma}(k_\perp \rho)e^{i(m+\sigma)\phi}e^{i(-\omega t+k_3 z)}.
	\end{equation}
	Although this ansatz method provides relatively simpler expressions for vector Bessel beams, the resulting mode functions fail to form an orthonormal basis. Indeed, the inner product (defined later in Eq. \eqref{eq.innerProduct}) between two modes $A^\mu_I(x^\alpha)$ and $A^\mu_{I'}(x^\alpha)$ is given by
	\begin{align}
		&\left <A^\mu_{I}(x^\alpha),A^\mu_{I'}(x^\alpha) \right >\nonumber \\ \propto& 2\omega \delta(k_3-k'_3)\delta (k_\perp-k'_\perp)\left [\frac 2 k_\perp \delta(m-m')\delta(\sigma-\sigma')\right .\nonumber \\& \left . +\frac{k_\perp \sigma\sigma'}{k^2_3}\delta(m+\sigma-m'-\sigma')\right ].
	\end{align}
	When $\sigma'=-\sigma$, the inner product is proportional to $-2\omega k_\perp\delta(k_3-k'_3)\delta (k_\perp-k'_\perp)\delta(m-m'+2\sigma)\delta(\sigma+\sigma')$, indicating that the modes $A^\mu_{k_3,k_\perp,\sigma, m}(x^\alpha)$ and $A^\mu_{k_3,k_\perp,-\sigma, m+2\sigma}(x^\alpha)$ exhibit crosstalk. Consequently, this method proves unsuitable for our investigation.
	
	Instead, we employ the Hertz potentials $\Theta_1$ and $\Theta_2$ \cite{10.1098/rspa.1955.0170,10.1098/rspa.1957.0092,PhysRevA.71.033411}, defined as
	\begin{equation}
		\Theta_i=C_i J_m(k_\perp \rho)e^{i(-\omega t+k_3 z+m\phi)},
	\end{equation}
	where $i \in \{1, 2\}$. The corresponding vector potential components are
	\begin{equation}
		A^t=-\Theta_{1,z},~~	A^\rho=\frac 1 \rho \Theta_{2,\phi},~~	A^\phi=-\Theta_{2,\rho},~~	A^z=\Theta_{1,t}.
	\end{equation}
	While this potential automatically satisfies the Lorenz gauge $A^\mu_{~~,\mu}=0$, residual gauge freedom remains. We explicitly impose the radiation gauge condition by requiring $A^t$ to transform to
	\begin{equation}
		A^t = -\Theta_{1,z}-\Lambda_{,t}=0,
	\end{equation}
	where $\Lambda$ is an arbitrary scalar function. Solving this relation yields
	\begin{equation}
		\Lambda=C_1 \frac{k_3}{\omega}J_m(k_\perp \rho)e^{i(-\omega t+k_3 z+m\phi)}=\frac{k_3}{\omega}\Theta_1.
	\end{equation}
	Concurrently, the spatial components of the vector potential transform as
	\begin{align}
		A^\rho=&\frac 1 \rho \Theta_{2,\phi}+\Lambda_{,\rho}=\frac 1 \rho \Theta_{2,\phi}+\frac {k_3}{\omega}\Theta_{1,\rho},\\A^\phi=&-\Theta_{2,\rho}+\frac 1 \rho \Lambda_{,\phi}=-\Theta_{2,\rho}+\frac {k_3}{\rho\omega}\Theta_{1,\phi},\\A^z=&\Theta_{1,t}+\Lambda_{,z}=\Theta_{1,t}+\frac {k_3}{\omega}\Theta_{1,z}.
	\end{align}
	From this transformed potential, we construct two independent sets of modes \cite{PhysRevA.71.033411}. Setting $C_1=0$ defines the transverse electric modes
	\begin{align}
		A^{{\rm (TE)}\hat \alpha}_{k_3k_\perp m}=&C_2\left [\frac {im}{\rho}J_m(k_\perp \rho)\hat \rho-k_\perp J^{(1)}_m(k_\perp \rho)\hat \phi \right ]\nonumber \\&\times e^{i(-\omega t+k_3 z+m \phi)},
	\end{align}
	where $J^{(1)}_m(\cdot)$ denotes the first derivative of the Bessel function with respect to its argument. Conversely, setting $C_2=0$ yields the transverse magnetic modes
	\begin{align}
		A^{{\rm (TM)}\hat \alpha}_{k_3k_\perp m}=&C_1\left [\frac{k_3k_\perp}{\omega}J^{(1)}_m(k_\perp \rho)\hat \rho+\frac{imk_3}{\rho\omega}J_m(k_\perp \rho)\hat \phi\right .\nonumber \\&\left . -\frac{ik^2_\perp}{\omega}J_m(k_\perp \rho)\hat z \right ]e^{i(-\omega t+k_3 z+m \phi)}.
	\end{align}
	As established previously, hatted indices (such as $\hat \alpha$) indicate an expansion within the cylindrical orthonormal basis $\{\hat t,\hat \rho,\hat \phi,\hat z \}$. Because these modes share an identical normalization requirement, we define a unified normalization factor $C_1=C_2=C$.
	
	Furthermore, we construct the circular polarization modes as $A^{\hat \alpha}_{k_3k_\perp\sigma m}=\left (A^{{\rm (TM)}\hat \alpha}_{k_3k_\perp m}-i\sigma A^{{\rm (TE)}\hat \alpha}_{k_3k_\perp m}\right )/\sqrt{2}$, where $\sigma=-1~ (1)$ represents right- (left-) circular polarization \cite{PhysRevA.71.033411,collett2005field,PhysRevLett.88.053601}, yielding\begin{widetext}
		\begin{equation}
			A^{(0)\hat \alpha}_{I}=\frac {Ce^{i(-\omega t+k_3 z+m \phi)}}{\sqrt 2}\Bigg ( \left [\frac{\sigma m}{\rho}J_m(k_\perp \rho)+ \frac{k_3k_\perp}{\omega}J^{(1)}_m(k_\perp \rho) \right ]\hat \rho+ i\left [\frac{mk_3}{\rho\omega}J_m(k_\perp \rho)+ \sigma k_\perp  J^{(1)}_m(k_\perp \rho) \right ]\hat \phi-\frac {ik^2_\perp}{\omega}J_m(k_\perp \rho)\hat z \Bigg ),\label{AinCylin}
		\end{equation}
		where the superscript $(0)$ indicates that the mode functions are established in Minkowski spacetime and $I$ represents the index set $\{k_3,k_\perp,\sigma, m\} $. We calculate the corresponding electric and magnetic fields as
		\begin{align}
			E^{(0)\hat \alpha}_{I}=&\frac {Ce^{i(-\omega t+k_3 z+m \phi)}}{\sqrt 2}\left (i \left [\frac{\omega\sigma m}{\rho}J_m(k_\perp \rho)+  k_3k_\perp J^{(1)}_m(k_\perp \rho) \right ]\hat \rho- \left [\frac{mk_3}{\rho}J_m(k_\perp \rho)+\omega \sigma k_\perp  J^{(1)}_m(k_\perp \rho) \right ]\hat \phi+k^2_\perp J_m(k_\perp \rho)\hat z \right ),\label{EinCylin}\\B^{(0)\hat \alpha}_{I}=&\frac {Ce^{i(-\omega t+k_3 z+m \phi)}}{\sqrt 2}\left ( \left [\frac{\omega m}{\rho}J_m(k_\perp \rho)+ \sigma k_3k_\perp J^{(1)}_m(k_\perp \rho) \right ]\hat \rho+i \left [\frac{\sigma mk_3}{\rho}J_m(k_\perp \rho)+ \omega  k_\perp  J^{(1)}_m(k_\perp \rho) \right ]\hat \phi-i \sigma k^2_\perp J_m(k_\perp \rho)\hat z \right ).\label{BinCylin}
		\end{align}
		Although expanding the fields \eqref{AinCylin} in the cylindrical basis provides an explicit phase factor $e^{im\phi}$ useful for analyzing OAM, these expressions become highly cumbersome during calculating spatial derivatives. To circumvent this difficulty, we transform the vector potential into the Cartesian basis, giving
		\begin{align}
			A^{(0)\alpha}_{I}=&\frac {Ce^{i(-\omega t+k_3 z+m \phi)}}{\sqrt 2 \omega}\Bigg ( \left [ \frac{m(\omega\sigma\cos\phi-ik_3\sin\phi)}{\rho}   J_m(k_\perp \rho)+     k_\perp (k_3\cos\phi-i\omega \sigma \sin\phi)J^{(1)}_m(k_\perp \rho) \right ]\hat x\nonumber \\& + \left [\frac{m(ik_3\cos\phi+\omega\sigma\sin\phi)}{\rho}J_m(k_\perp \rho)+ k_\perp(i\omega\sigma\cos\phi+k_3\sin\phi)  J^{(1)}_m(k_\perp \rho) \right ]\hat y-ik^2_\perp J_m(k_\perp \rho)\hat z \Bigg ).\label{AinCartesian}
		\end{align}
	\end{widetext}
	Because this Cartesian representation contains trigonometric functions of $\phi$, it obscures the explicit spatial phase structure. Also, as demonstrated later, the parameter $m$ does not represent the OAM, despite the presence of the $e^{im\phi}$ term. Rather, $m$ denotes the total AM of the field, while $l=m-\sigma$ corresponds to the OAM within the paraxial approximation. Last, we express a general EM field $A^{\alpha}$ in Minkowski spacetime through the standard mode expansion
	\begin{equation}
		A^{\alpha}(x^\mu)=A^{(0)\alpha}(x^\mu)=\sum_{I}\left (A^{(0)\alpha}_{I}a_{I}+A^{(0)\alpha *}_{I}a^*_{I}\right ),
	\end{equation}
	where the asterisk ($*$) denotes complex conjugation, $a_{I}$ are the mode coefficients, and the summation symbol $\sum_I$ represents $\sum_{\sigma}\sum_{m}\int dk_3\int dk_\perp$.
	
	Following the canonical quantization procedure of quantum field theory \cite{peskin2018introduction,scully1997quantum}, the EM field in Minkowski spacetime takes the quantized form
	\begin{equation}
		\hat A^{\alpha}_{\rm MS}(x^\mu)=\sum_{I}\left (A^{(0)\alpha}_{I}\hat a_{I}+A^{(0)\alpha *}_{I} \hat a^\dagger_{I}\right ),\label{eq.quantumFieldInMinkowski}
	\end{equation}
	where the classical coefficients $a_{I}$ are promoted to the quantum operators $\hat a_{I}=\hat a_{k_3k_\perp\sigma m}$. Similarly, the electric and magnetic fields are quantized as $\hat E^{\alpha}_{\rm MS}(x^\mu)=\sum_{I}\left (E^{(0)\alpha}_{I}\hat a_{I}+E^{(0)\alpha *}_{I} \hat a^\dagger_{I}\right )$ and $\hat B^{\alpha}_{\rm MS}(x^\mu)=\sum_{I}\left (B^{(0)\alpha}_{I}\hat a_{I}+B^{(0)\alpha *}_{I} \hat a^\dagger_{I}\right )$. For the sake of notational simplicity, we drop the hats on these operators hereafter where no confusion arises. The Hamiltonian of the EM field is expressed as
	\begin{align}
		H=&:\frac{\epsilon_0}2 \int d^3x^i \left (E^\mu E^\dagger_\mu +B^\mu B^\dagger_\mu\right ):\nonumber \\=&\frac {\epsilon_0} 2 \sum_I \sum_{I'}\int d^3x^i \Big [(E^\mu_I E_{I' \mu}+B^\mu_I B_{I' \mu})a_{I}a_{I'}+(E^{\mu*}_I E^*_{I' \mu}\nonumber \\&+B^{\mu*}_I B^*_{I' \mu})a^\dagger_{I}a^\dagger_{I'}+2(E^{\mu}_I E^*_{I' \mu}+B^{\mu}_I B^*_{I' \mu})a^\dagger_{I}a_{I'} \Big ]\nonumber \\=&\sum_I 8 \epsilon_0 \pi^2 C^2 k_\perp \omega^2 a^\dagger_I a_I,
	\end{align}
	where $\epsilon_0$ is the vacuum permittivity, and terms proportional to $a_Ia_{I'}$ and $a_I^\dagger a_{I'}^\dagger$ are neglected due to their rapid $2\omega$ oscillations. Assuming the measurement timescale is much greater than $(2\omega)^{-1}$, these highly oscillatory terms average to zero and yield no observable effect. The symbol $:~:$ represents normal ordering where all creation operators are placed to the left of all annihilation operators in products \cite{peskin2018introduction}. By imposing the energy requirement $H=:\sum_I \omega(a^\dagger_I a_I+a_I a^\dagger_I)/2:=\sum_I \omega a^\dagger_I a_I$, we uniquely determine the normalization factor to be $C=(8 \epsilon_0 \pi^2  k_\perp \omega)^{-1/2}$.
	
	The SAM of the quantum EM field is defined as \cite{PhysRevResearch.4.023165,barnett2016natures}
	\begin{align}
		{\mathbf s}=&\epsilon_0 \int d^3 x^i : {\mathbf  E}_\perp \times  {\mathbf  A}_\perp :\nonumber \\=&-\epsilon_0\int d^3x^i :\sum_{I}\left (-i\omega { \mathbf A}^{(0)}_{I\perp} a_{I}+i\omega {\mathbf A}^{(0) *}_{I\perp}  a^\dagger_{I}\right )\nonumber \\ &\times\sum_{I'}\left ({\mathbf A}^{(0)}_{I'\perp} a_{I'}+{\mathbf A}^{(0) *}_{I'\perp}  a^\dagger_{I'}\right ):\nonumber \\=&\epsilon_0 \sum_{I} \sum_{I'} i\omega\int d^3x^i \left (  { \mathbf A}^{(0)}_{I\perp}\times {\mathbf A}^{(0) *}_{I'\perp}  a^\dagger_{I'}a_{I}\right .\nonumber \\&\left .- {\mathbf A}^{(0) *}_{I\perp}\times {\mathbf A}^{(0)}_{I'\perp}   a^\dagger_{I}  a_{I'}\right )\nonumber \\=&\sum_I \frac {k_3}{\omega} \sigma a^\dagger_I a_I \hat z,
	\end{align}
	where the highly oscillatory terms proportional to $a_Ia_{I'}$ and $a_I^\dagger a_{I'}^\dagger$ are again neglected. The subscript $\perp$ denotes the transverse components, such that ${ \mathbf A}^{(0)}_{I\perp}=A^{(0)1}_{I}\hat x+A^{(0)2}_{I}\hat y$. While the SAM for standard plane waves equals $\sigma$, it is scaled by the geometric factor $k_3/\omega$ for vortex beams. Consequently, a right- (left-) handed circular polarization state with $\sigma=-1$ ($1$) possesses a SAM of $- k_3/\omega$ ($ k_3/\omega$) along the propagation direction. Under the paraxial approximation where $k_3/\omega\rightarrow 1$, this SAM naturally reduces to $\sigma$. Furthermore, the OAM operator evaluated about an arbitrary point $\mathbf r_0$ is expressed as \cite{PhysRevA.71.033411,PhysRevResearch.4.023165}
	\begin{equation}
		\mathbf L (\mathbf r_0)=:\epsilon_0 \int d^3x^iE_{\perp\mu}[(\mathbf x-\mathbf r_0)\times \nabla]A^{(0)\mu}_\perp:.
	\end{equation}
	Setting the spatial origin to $\mathbf r_0=0$ yields
	\begin{align}
		\mathbf L=&\mathbf L (0)=:\epsilon_0 \int d^3x^iE_{\perp\mu}(\mathbf x\times \nabla)A^{(0)\mu}_\perp:\nonumber \\=&\epsilon_0\sum_{II'}i\omega\int d^3x^i(A^{(0)}_{I\perp\mu}\boldsymbol{\mathcal L} A^{(0)\mu *}_{I'\perp}  a^\dagger_{I'}a_{I}-A^{(0)*}_{I\perp\mu}\boldsymbol{\mathcal L} A^{(0)\mu }_{I'\perp}a^\dagger_{I}  a_{I'}),
	\end{align}
	with the differential operator $\boldsymbol{\mathcal L}=\hat x(-z\sin\phi\partial_\rho-z\rho^{-1}\cos\phi\partial_\phi+\rho\sin\phi\partial_z)+\hat y(z\cos\phi\partial_\rho-z\rho^{-1}\sin\phi\partial_\phi-\rho\cos\phi\partial_z)+\hat z\partial_\phi$. Unlike the Cartesian basis, the azimuthal derivatives of the cylindrical basis vectors do not vanish, satisfying $\partial_\phi \hat \rho=\hat \phi$ and $\partial_\phi \hat \phi=-\hat \rho$. The longitudinal component of the OAM operator therefore evaluates to
	\begin{equation}
		L_3=\sum_I \left (m-\frac {k_3}{\omega}\sigma \right ) a^\dagger_I a_I=\sum_I l a^\dagger_I a_I,
	\end{equation}
	where $l=m-k_3\sigma/\omega$ denotes the OAM of the mode. The secondary term $-k_3\sigma/\omega$ emerges from the spatial derivatives of the $\hat \rho$ and $\hat \phi$ basis vectors. Because determining the OAM involves measuring the azimuthal phase variation, the inherent continuous rotation of these cylindrical basis vectors simultaneously encodes the SAM information. In total, a single photon occupying the $A^{(0)}_{k_3k_\perp\sigma m}$ mode carries a SAM of $\sigma k_3/\omega $ ($\approx \sigma$ in the paraxial approximation) and an OAM of $l=m-k_3\sigma/\omega$ ($\approx m-\sigma$ in the paraxial approximation) \cite{PhysRevA.82.063825,bliokh2015spin}.
	
	\section{Vector Bessel Light fields in GWs}
	
	In linearized gravity, the spacetime metric for a monochromatic GW propagating along the $z$ axis is expressed perturbatively as
	\begin{equation}
		g_{\alpha\beta}(x^\mu)=\eta_{\alpha\beta}+\epsilon h_{\alpha\beta}(x^\mu),
	\end{equation}
	where $\eta_{\alpha\beta}$ is the Minkowski metric, $h_{\alpha\beta}(x^\mu)$ represents the metric perturbation induced by the GWs, and $\epsilon$ serves as a bookkeeping parameter to track the perturbation order. Setting $\epsilon\rightarrow 1$ at the end of the calculation recovers the full physical metric perturbation \cite{sakurai2020modern,exirifard2021towards}. Imposing the transverse-traceless (TT) gauge, the metric perturbation takes the form
	\begin{equation}
		h_{\alpha\beta}(x^\mu)=A_\lambda e^{ik_g(-t+z)} e^\lambda_{\alpha\beta},
	\end{equation}
	where the index $\lambda=\{+,\times\}$ denotes the GW polarization states, $A_\lambda$ represents the corresponding amplitude, $k_g$ defines the GW frequency, and $e^\lambda_{\alpha\beta}$ are the unit linear polarization tensors \cite{misner1973gravitation}. For mathematical convenience, we absorb the amplitudes into a combined polarization tensor defined by
	\begin{equation}
		\tilde \epsilon_{\alpha\beta}=A_\lambda  e^\lambda_{\alpha\beta}=\begin{pmatrix} 0&0&0&0\\ 0&A_+&A_\times&0\\ 0&A_\times&-A_+&0\\ 0&0&0&0 \end{pmatrix}.
	\end{equation}
	
	Without loss of generality, we consider a monochromatic light field propagating along the $z$ axis that begins interacting with a monochromatic GW at time $t=0$. Given that the vortex beam exhibits axial symmetry around its propagation axis \cite{andrews2021symmetry}, we can, without loss of generality, rotate the coordinates to allow the GW to propagate at an angle $\theta$ relative to the $z$ axis within the $xz$ plane. The corresponding metric perturbation $h_{\alpha\beta}(x^\mu)$ takes the form
	\begin{equation}
		h_{\alpha\beta}(x^\mu)=\epsilon_{\alpha\beta}  e^{ik_g(-t+x\sin\theta +z\cos\theta )},
	\end{equation}
	where the transformed polarization tensor $\epsilon_{\alpha\beta}$ relates to the original tensor $\tilde \epsilon_{\alpha\beta}$ through the transformation
	\begin{equation}
		\epsilon_{\alpha\beta}=(R_y)^{\mu}_{\alpha} \tilde \epsilon_{\mu\nu} (R^{-1}_y)^{\nu}_{\beta} ,
	\end{equation}
	with $R_y$ representing the rotation matrix about the $y$ axis and $R^{-1}_y$ denoting its inverse.
	
	The action governing the EM field is \cite{dewitt1975quantum,peskin2018introduction}
	\begin{equation}
		S[A_\mu]=\int d^4 x^\mu \sqrt{-g} \frac {-1}{4}F_{\mu\nu}F^{\mu\nu}=\int d^4 x^\mu \mathcal L,
	\end{equation}
	where $F_{\mu\nu}=A_{\nu;\mu}-A_{\mu;\nu}$ defines the EM tensor and $\mathcal L$ represents the Lagrangian density. Varying this action with respect to $A_\mu$ yields the equations of motion (EOMs) for the EM field
	\begin{equation}
		A^{\mu;\nu}_{~~~;\nu}-A^{\nu~;\mu}_{~;\nu}-R^\mu_{~\alpha}A^\alpha=0,
	\end{equation}
	where $R^\mu_{~\alpha}=g^{\mu\rho}g^{\nu\beta}R_{\beta\alpha\nu\rho}$ denotes the Ricci tensor associated with the metric $g_{\mu\nu}$. To resolve the two gauge degrees of freedom inherent in the EM field, we impose the radiation gauge conditions
	\begin{equation}
		A^0=A^i_{~;i}=A^\mu_{~;\mu}=0.
	\end{equation}
	Following this gauge fixing, the EOMs simplify to
	\begin{equation}
		A^{\mu;\nu}_{~~~;\nu}-R^\mu_{~\alpha}A^\alpha=0.
	\end{equation}
	For the GW, our previous selection of the TT gauge dictates
	\begin{equation}
		h=h^\alpha_{~\alpha}=h_{0\mu}=h^{\mu\nu}_{~~,\nu}=0.
	\end{equation}
	Additionally, the metric perturbation $h_{\mu\nu}$ satisfies the wave equation
	\begin{equation}
		h^{\alpha\beta,\mu}_{~~~~~,\mu}=0.
	\end{equation}
	Consequently, the Ricci tensor evaluates to
	\begin{align}
		R^\mu_{~\alpha}=&\frac \epsilon 2 \left (h^{\beta~,\mu}_{~\nu~~,\beta}+h^{\mu\alpha}_{~~,\alpha\nu}-h^{\beta~,\mu}_{~\beta~~,\nu}-h^{\mu~,\beta}_{~\nu~~,\beta} \right )+O(\epsilon^2)\nonumber \\=&O(\epsilon^2),
	\end{align}
	and the EOMs expand to first order in $\epsilon$ as
	\begin{equation}
		A^{\mu,\nu}_{~~~,\nu}-\epsilon [h^{\alpha\beta}A^\mu_{~,\alpha\beta}-(h^{\mu\beta}_{~~,\alpha}-h^{\mu~,\beta}_{~\alpha}-h^{\beta~,\mu}_{~\alpha})A^\alpha_{~,\beta}]=0.\label{eq.EOMs}
	\end{equation}
	Because GW amplitudes are exceedingly small, we expand the EM field perturbatively as
	\begin{equation}
		A^\alpha=A^{(0)\alpha}+\epsilon A^{(1)\alpha},
	\end{equation}
	where $A^{(0)\alpha}=A^{(0)\alpha}(x^\mu)$ represents the unperturbed EM field and $A^{(1)\alpha}=A^{(1)\alpha}(x^\mu)$ denotes the perturbation induced by the GW. Substituting this expansion for $A^\alpha$ into Eq. \eqref{eq.EOMs} separates the dynamics into the zeroth order and first order EOMs,
	\begin{align}
		A^{(0)\mu,\nu}_{~~~~~~,\nu}=&0,\label{eq.zerothEOM}\\A^{(1)\mu,\nu}_{~~~~~~,\nu}=&h^{\alpha\beta}A^{(0)\mu}_{~~~~,\alpha\beta}+(h^{\beta~,\mu}_{~\alpha}-h^{\mu\beta}_{~~,\alpha}-h^{\mu~,\beta}_{~\alpha})A^{(0)\alpha}_{~~~~,\beta}.\label{eq.firstEOM}
	\end{align}
	
	To evaluate the derivatives on the right-hand side of Eq. \eqref{eq.firstEOM}, $A^{(0)\mu}$ must first be expanded in a Cartesian basis, as noted previously. Otherwise, it requires differentiating the cylindrical basis vectors, which is unnecessarily cumbersome. Because the components of $A^{(0)\mu}$ are expressed in cylindrical coordinates, the Cartesian derivatives must be transformed to cylindrical coordinates via $\partial_\alpha=\Lambda^{\tilde \mu}_{~~\alpha}\partial_{\tilde \mu}$, where
	\begin{equation}
		\Lambda^{\tilde \mu}_{~~\alpha}=\begin{pmatrix} 1&0&0&0\\ 0&\cos\phi&\sin\phi&0\\ 0&-\rho^{-1}\sin\phi&\rho^{-1}\cos\phi&0\\ 0&0&0&1 \end{pmatrix}.
	\end{equation}
	Moreover, the cylindrical coordinate basis is unnormalized, but the EM field should be expanded within an orthonormal basis. The transformation mapping the perturbed EM field from the Cartesian basis to the cylindrical orthonormal basis takes the form $A^{(1)\hat \alpha}=T^{\hat \alpha}_{~~\beta}A^{(1)\beta}$ where
	\begin{equation}
		T^{\hat \alpha}_{~~\beta}=\begin{pmatrix} 1&0&0&0\\ 0&\cos\phi&\sin\phi&0\\ 0&-\sin\phi&\cos\phi&0\\ 0&0&0&1 \end{pmatrix}.
	\end{equation}
	
	Furthermore, we solve the EOMs \eqref{eq.firstEOM} by using the Green's function method \cite{q38s-k2tq}. A general wave equation with a source term $F(x^\mu)$,
	\begin{equation}
		(-\partial^2_0+\partial^i\partial_i)\psi(x^\mu)=F(x^\mu),\label{eq.waveEquation}
	\end{equation}
	admits a solution of the form \cite{duffy2015green}
	\begin{equation}
		\psi(x^\mu)=\int d^4 x^{\mu'}\mathcal G(x^\mu,x^{\mu'}) F(x^{\mu'}),\label{eq.greenFun}
	\end{equation}
	where the $3+1$ dimensional Green's function expressed in cylindrical coordinates is \cite{q38s-k2tq}
	\begin{align}
		\mathcal G(x^\mu,x^{\mu'})=&\frac{1}{8\pi^3}\sum_{n=-\infty}^\infty \int_{-\infty}^\infty dp_3 \int_0^\infty d p_\perp\int dp_0 \frac{p_\perp }{p^2_0-E^2}\nonumber \\ &\times  e^{-ip_0(t-t')+ip_3(z-z')+in(\phi-\phi')} J_n(p_\perp \rho) \nonumber \\ &\times J_n(p_\perp \rho')\theta(t-t'),\label{eq.GreenFunctionIntegral} 
	\end{align}
	with $E=(p_3^2+p_\perp^2)^{1/2}$. The integral with respect to $p_0$ represents a contour integration in the complex plane from $-\infty$ to $\infty$ that is closed in the lower half plane.
	
	We adopt the sandwich model \cite{carneiro2025gravitational,chakraborty2024particle,gibbons1975quantized} where the light field interacts with the GWs strictly within the time interval $[0, t]$. By comparing Eq. \eqref{eq.firstEOM} and Eq. \eqref{eq.waveEquation}, we identify the source term $F^\nu\left [\mathbf{A}^{(0)}(x^{\mu}) \right ]$ as
	\begin{align}
		&F^\nu\left [\mathbf{A}^{(0)}(x^{\mu}) \right ]\nonumber \\ =&h^{\alpha\beta}(x^\mu)A^{\nu}_{~,\alpha\beta}(x^\mu)+\Big [h^{\beta~,\nu}_{~\alpha}(x^\mu)-h^{\nu\beta}_{~~,\alpha}(x^\mu)-h^{\nu~,\beta}_{~\alpha}(x^\mu)\Big ]\nonumber \\ &\times A^{\alpha}_{~,\beta}(x^\mu)\nonumber \\ =&h^{\alpha\beta}(x^\mu)\Lambda^{\tilde \gamma}_{~~\alpha} \Lambda^{\tilde \lambda}_{~~\beta}  A^{\nu}_{~,\tilde \gamma \tilde \lambda}(x^\mu)+\eta^{\beta\gamma}\eta^{\nu\delta}\Big [h_{\gamma\alpha,\delta}(x^\mu)\nonumber \\ &-h_{\delta\gamma,\alpha}(x^\mu)-h_{\delta\alpha,\gamma}(x^\mu)\Big ] \Lambda^{\tilde \lambda}_{~~\beta}  A^{\alpha}_{~,\tilde \lambda}(x^\mu). 
	\end{align}
	The solution for the perturbed field $A^{(1)\hat \sigma}$ in the cylindrical orthonormal basis is then given by
	\begin{align}
		&A^{(1)\hat \sigma}(x^\mu)=T^{\hat \sigma}_{~~\nu}A^{(1)\nu}(x^\mu)\nonumber \\ =&T^{\hat \sigma}_{~~\nu}\int d^4 x^{\mu'}\mathcal G(x^\mu,x^{\mu'}) \theta(t')F^\nu\left [\mathbf{A}^{(0)}(x^{\mu'}) \right ]\nonumber \\ =& T^{\hat \sigma}_{~~\nu}\int d^4 x^{\mu'}\mathcal G(x^\mu,x^{\mu'}) \theta(t')\sum_I\Big (F^\nu\left [\mathbf{A}^{(0)}_I(x^{\mu'}) \right ]a_I\nonumber \\ &+F^\nu\left [\mathbf{A}^{(0)*}_I(x^{\mu'}) \right ]a^*_I\Big )\nonumber \\ =&\sum_{I}\Big (A^{(1)\hat \sigma}_{I}a_{I}+A^{(1)\hat \sigma *}_{I}a^*_{I}\Big ), 
	\end{align}
	where the Heaviside function $\theta(t')$ restricts the interaction time to $t'>0$. The perturbation mode function $A^{(1)\hat \sigma}_{I}(x^\mu)$ is expressed as
	\begin{align}
		A^{(1)\hat \sigma}_{I}=&T^{\hat \sigma}_{~~\nu}\int d^4 x^{\mu'}\mathcal G(x^\mu,x^{\mu'}) \theta(t')F^\nu\left [\mathbf{A}^{(0)}_I(x^{\mu'}) \right ]\nonumber \\ =&\int  dp_3  dp_\perp dp_0   d\rho'   dz' dt'\frac{\rho'p_\perp}{8\pi^3(p^2_0-E^2)}\theta(t')\theta(t-t')\nonumber \\ &\times e^{-ip_0(t-t')+ip_3(z-z')}\mathcal I_I^{\hat\sigma},\label{eq.perturbationModes}
	\end{align}
	with the angular integral
	\begin{equation}
		\mathcal {I}_I^{\hat\sigma}=\sum_n\int d\phi'e^{in(\phi-\phi')}J_n(p_\perp \rho)J_n(p_\perp \rho')T^{\hat \sigma}_{~~\nu}F^\nu\left [\mathbf{A}^{(0)}_I(x^{\mu'}) \right ].
	\end{equation}
	
	Taking the component $\mathcal {I}_I^{\hat\rho}$ as an example, we have
	\begin{align}
		&\mathcal {I}_I^{\hat\rho}\nonumber \\ =&\sum_n\int d\phi' \frac {Ce^{i[-(\omega+k_g)t'+(k_3+k_g\cos\theta)z'+k_g\sin\theta x'+n\phi]}}{16\sqrt 2\omega}J_n(p_\perp \rho)\nonumber \\ &\times J_n(p_\perp \rho')e^{i(m-n)\phi'}\sum_{j=-3}^3 e^{ij\phi'}(e^{-i\phi}T_{j,1}+e^{i\phi}T_{j,2})\nonumber \\ =&\frac {\pi Ce^{i[-(\omega+k_g)t'+(k_3+k_g\cos\theta)z'+k_g\sin\theta x']}}{8\sqrt 2 \omega}\sum_{j=-3}^3  J_{m+j}(p_\perp \rho)\nonumber \\ &\times J_{m+j}(p_\perp \rho') [e^{i(m+j-1)\phi}T_{j,1}+e^{i(m+j+1)\phi}T_{j,2}]  \nonumber \\ =&\frac {\pi Ce^{i[-(\omega+k_g)t'+(k_3+k_g\cos\theta)z'+k_g\sin\theta x']}}{8\sqrt 2 \omega}\sum_{j=-3}^3 e^{i(m+j)\phi}S_j. 
	\end{align}
	The terms $T_{j,1}$ and $T_{j,2}$ are deduced from the definition of $\mathcal {I}_I^{\hat\rho}$. We find that $T_{3,2}=T_{-3,1}=0$, and the linear superposition of $T_{j,1}$ and $T_{j,2}$ can be rearranged into terms with well-defined OAM expressions $e^{i(m+j)\phi}S_j$ where
	\begin{align}
		S_j=&J_{m+j-1}(p_\perp \rho)J_{m+j-1}(p_\perp \rho') T_{j-1,2}\nonumber \\
		&+J_{m+j+1}(p_\perp \rho)J_{m+j+1}(p_\perp \rho')T_{j+1,1},
	\end{align}
	with $T_{4,1}=T_{-4,2}=0$. The terms $S_{\pm 2}$ and $S_0$ are more complex. For example, $S_2$ scales as
	\begin{equation}
		S_2\propto \sum_{a,b,n} C_{a,b,n} \rho'^n J_{m+2}^{(a)}(k_\perp \rho')J_{m+2}^{(b)}(p_\perp \rho'),
	\end{equation}
	where all Bessel functions are transformed into $(m+2)$th order, the indices $a$ and $b$ take values of 0 or 1, and $C_{a,b,n}$ are constants with $n\le 0$. During the subsequent integration over $\rho'$, we observe that in the limit $\rho' \rightarrow 0$ (specifically the region $\rho' \lesssim k_\perp^{-1}$), the Bessel function behaves as $J_m(x)\sim [2\theta(m)-1]^m (x/2)^{|m|}/(|m|!)$. Consequently, $S_2 \sim D+\mathcal O((\rho')^0)$ for some constant $D$. This region contributes insignificantly to the total integral and is safely neglected. In the remaining radial domain, since $n\le 0$, we retain only the leading order terms in $\rho'$. After calculating $S_j$, we substitute $\mathcal {I}_I^{\hat\rho}$ into Eq. \eqref{eq.perturbationModes} to evaluate $A^{(1)\hat \rho}_{I}$. The integrations over $p_3$ and $z'$ yield
	\begin{equation}
		\int dp_3dz'e^{i[p_3(z-z')+(k_3+k_g\cos\theta)z']}=2\pi e^{i(k_3+k_g\cos\theta)z},
	\end{equation}
	while the temporal integrations over $p_0$ and $t'$ result in
	\begin{align}
		&\int dp_0dt' \frac{\theta(t')\theta(t-t')e^{-ip_0(t-t')-i(\omega+k_g)t'}}{(p^2_0-E^2)} \nonumber \\ \approx&\frac{\pi[e^{-iEt}-e^{-i(\omega+k_g)t}]}{E(E-\omega-k_g)},
	\end{align}
	where we neglect the term proportional to $1/(E+\omega+k_g)$ and define $E=[(k_3+k_g\cos\theta)^2+k^2_\perp]^{1/2}\approx\omega+k_3k_g\cos\theta/\omega$. The remaining integrations over $p_\perp$ and $\rho'$ are performed in two parts. The first involves integrating over $\rho'$ to produce $\delta(k_\perp-p_\perp)$ followed by an integration over $p_\perp$. The second involves integrating over $p_\perp$ to yield $\delta(\rho-\rho')$ followed by an integration over $\rho'$. As shown subsequently, the perturbation modes comprise seven different OAM modes. Due to this complexity, these explicit integration results are omitted here. An analogous technique is used to calculate $A^{(1)\hat \phi}_{I}$ and $A^{(1)\hat z}_{I}$.
	
	Analogous to the quantization procedure in Minkowski spacetime, the EM field in the presence of GWs is quantized as \cite{birrell1984quantum,chakraborty2024particle}
	\begin{equation}
		\hat A^{\alpha}_{\rm GW}(x^\mu)=\sum_{I}\left (A^{\alpha}_{I} \hat b_{I}+A^{\alpha *}_{I} \hat b^\dagger_{I}\right ),\label{eq.quantumFieldInGW}
	\end{equation}
	where $A^{\alpha}_{I}=A^{(0)\alpha}_{I}+\epsilon A^{(1)\alpha}_{I}$ denotes the set of mode functions within the GW. In this representation, $\hat b_{I}$ and $\hat b^\dagger_{I}$ represent the annihilation and creation operators corresponding to the perturbed mode $A^{\alpha}_{I}$.
	
	\section{Quantum states of vector Light fields in GWs}
	
	The Bogoliubov transformation provides the connection between the fields defined in Eq. \eqref{eq.quantumFieldInMinkowski} and Eq. \eqref{eq.quantumFieldInGW}. The resulting relationship between the operators $\hat b_{I}$ and $\hat a_{I}$ is expressed as \cite{chakraborty2024particle}
	\begin{equation}
		\hat b_{I}= \sum_{I'} \left (\alpha^*_{I,I'}\hat a_{I'}-\beta^*_{I,I'}\hat a^\dagger_{I'} \right ),
	\end{equation}
	where $\alpha_{I,I'}=\left <A^{(0)\mu}_{I'} ,A^{\mu}_{I}\right >$ and $\beta_{I,I'}=-\left <A^{(0)\mu*}_{I'} ,A^{\mu}_{I}\right >$ are the Bogoliubov coefficients and $\left <\cdot ,\cdot \right >$ represents the Klein-Gordon inner product. The coefficients $\alpha_{I,I'}$ represent the projection of the perturbed light field $A^{\mu}_{I}$ onto the orthonormal Minkowski basis $A^{(0)\mu}_{I'}$. We note that the coefficients $\beta_{I,I'}$ are associated with photon creation induced by the GW \cite{chakraborty2024particle}. Because this represents an extremely weak physical process, we neglect $\beta_{I,I'}$ in the subsequent analysis.
	
	We adopt the Klein-Gordon inner product for a vector field $A^\alpha$ in curved spacetime \cite{PhysRevD.106.065002}. First, we define a current as
	\begin{equation}
		J^\mu[A^\alpha,A'^\alpha]=\frac{i}{\sqrt{-g}}\left (A^*_\nu \pi'^{\mu\nu}-\pi^{\mu\nu*}A'_\nu \right ),
	\end{equation}
	where $\pi^{\mu\nu}=\delta \mathcal L/(\delta A_{\nu;\mu})=\sqrt{-g}(A^{\mu;\nu}-A^{\nu;\mu})$. Because this current $J^\mu[A^\alpha,A'^\alpha]$ is divergence free, satisfying $J^\mu[A^\alpha,A'^\alpha]_{;\mu}=0$, it represents a conserved quantity. The Klein-Gordon inner product is then defined as \cite{takagi1986vacuum,PhysRevD.106.065002}
	\begin{equation}
		\left < A^\alpha,A'^\alpha\right >=N \int_\Sigma d^3 \mathbf x  n^\mu \sqrt{\gamma} J_\mu[A^\alpha,A'^\alpha],
	\end{equation}
	where $N$ is a constant, $\Sigma$ is a spacelike hypersurface for integration, $\gamma_{\mu\nu}$ is the induced 3-metric on this hypersurface with $\gamma=\det( \gamma_{\mu\nu})$, and $n^\mu$ is the unit normal vector. In the presence of GWs, we foliate the spacetime into a family of spacelike hypersurfaces parametrized by $t$. To first order in $\epsilon$, we have $n^\mu=(1,0,0,0)$ and $\gamma =1$. The inner product thus becomes
	\begin{align}
		&\left < A^\alpha,A'^\alpha\right >\nonumber \\=&N \int_\Sigma d^3 \mathbf x  i\left [A^{\nu*}\left (A'_{0,\nu}-A'_{\nu,0}\right )-\left (A^*_{0,\nu}-A^*_{\nu,0}\right )A^{'\nu}\right ]\nonumber \\ =&N \int_\Sigma d^3 \mathbf x  ig_{\mu\nu}\left (A^{\mu*}_{~~,0}A'^\nu-A^{\nu*}A'^\mu_{~~,0} \right ),\label{eq.innerProduct}
	\end{align}
	where the radiation gauge conditions have been applied in the final expression. When applying this inner product to the EM modes in the Minkowski metric, we find $\left <A^{(0)\alpha}_I ,A^{(0)\alpha}_{I'}\right >=- N \delta(I-I')/\epsilon_0$, with $\delta(I-I')=\delta(k_3-k'_3)\delta(k_\perp-k'_\perp)\delta(m-m')\delta(\sigma-\sigma')$. Since the mode functions $A^{(0)\alpha}_{I}$ are normalized such that $\left <A^{(0)\alpha}_I ,A^{(0)\alpha}_{I'}\right >=\delta(I-I')$, we can identify the constant $N=-\epsilon_0$.
	
	In the sandwich model \cite{carneiro2025gravitational,chakraborty2024particle,gibbons1975quantized}, the light field interacts with GWs within the time interval $[0, t]$ and the photon state is subsequently measured in the absence of GWs. Consequently, we calculate the inner product using the  metric $g_{\mu\nu}=\eta_{\mu\nu}$. The Bogoliubov coefficients $\alpha_{I,I'}$ are then evaluated as
	\begin{equation}
		\alpha_{I,I'}=\left <A^{(0)\mu}_{I'} ,A^{\mu}_{I} \right >=\delta(I-I')+\epsilon\left <A^{(0)\mu}_{I'} , A^{(1)\mu}_{I} \right >,
	\end{equation}
	where the inner product between the unperturbed field and the first order perturbation is expressed as\begin{widetext}
		\begin{equation}
			\left <A^{(0)\mu}_{I'} , A^{(1)\mu}_{I} \right >=\epsilon_0 \int_\Sigma d^3 \mathbf x  i\eta_{\mu\nu}\left (A^{(0)\nu*}_{I'}A'^{(1)\mu}_{I~~,0}-A^{(0)\mu*}_{I'~~,0}A'^{(1)\nu}_I \right )=\int_\Sigma d\rho d\phi dz g(x^\mu),
		\end{equation}	
		with the integrand $g(x^\mu)$ given by
		\begin{align}
			g(x^\mu)=&\frac {i\epsilon_0 C C'}{64E(E-k_g \omega)\omega}\left (\left [-iE e^{-iEt}+i(\omega+k_g)e^{-i (\omega +k_g)t} \right ]e^{i\omega' t}-i\omega'e^{i\omega' t}\left [e^{-iEt}-e^{-i(\omega+k_g)t} \right ] \right )e^{i(k_3+k_g\cos\theta-k'_3)z}\nonumber \\ &\times \left [\sum_{a,b,j}T^{(a,b)}_je^{i(m-m'+j)\phi}J^{(a)}_{m+j}(k_\perp \rho)J^{(b)}_{m'}(k'_\perp \rho)\rho^{a+b-1}+\sum_j R_j e^{i(m-m'+j)\phi}J_{m+j}(k_\perp \rho)J_{m'}(k'_\perp \rho)\rho \right ], 
		\end{align}
		where $j \in \{-3, \dots, 3\}$, $a, b \in \{0, 1\}$, and $J^{(0)}_{m}(x)=J_{m}(x)$. The normalization $C'=(8\epsilon_0\pi^2k'_\perp\omega')^{-1/2}$ and frequency $\omega'=(k'^2_3+k'^2_\perp)^{1/2}$ are used alongside the matrices $T_j$ and vector $R$ deduced from the inner product expression. To the lowest order of $k_g/k_3$, the inner product simplifies to
		\begin{equation}
			\left <A^{(0)\mu}_{I'} , A^{(1)\mu}_{I} \right > =f(t)\delta(k_3+k_g\cos\theta-k'_3)\delta (k_\perp-k'_\perp) \sum_{j=-3}^3\Big [C_j\delta(\sigma-\sigma')\delta(m+j-m')+D_{\sigma,j}\delta(\sigma+\sigma') \delta(m+j-m')\Big ], \label{eq.innerProBetween01}
		\end{equation}
		where $f(t)=[1-e^{ik_g t(\Theta^{-1}\cos\theta-1)}]/[64 \Theta (\cos\theta-\Theta)]$ and $\Theta=\sqrt{1+\gamma^2}=\omega/k_3$ with the paraxial parameter $\gamma=k_\perp/k_3$. We further evaluate the coefficients $C_j$ and $D_{\sigma,j}$ as
		\begin{align}
			C_{\pm 3}=&\gamma^3 [4A_\times \cos\theta\pm i A_+(3+\cos 2\theta)]\sin\theta/\Theta,\\
			C_{\pm 2}=&4\Theta \gamma^2k_3[\mp 4i A_\times \cos\theta+A_+(3+\cos2\theta)]/k_g,\\
			C_{\pm 1}=&32 \Theta \gamma k_3(A_\times \pm iA_+\cos\theta)\sin\theta/k_g,\\
			C_0=&16\Theta(\gamma^2-2)k_3A_+\sin^2\theta/k_g,\\
			D_{\sigma,\pm 3}=&i\gamma(\Theta\mp \sigma)^2[4iA_\times \cos\theta\mp A_+(3+\cos2\theta)]\sin\theta/\Theta,\\
			D_{\sigma,\pm 2}=&(\Theta\mp \sigma)^2\big (8A_\times\cos\theta[i\sigma(\Theta\pm \sigma)\cos\theta-\sigma(\sigma \pm \Theta)\sin\theta \mp i\Theta]+A_+[6\Theta-7\sigma(\sigma \pm \Theta)\cos\theta+2\Theta \cos2\theta\nonumber \\&+\sigma(\mp \Theta-\sigma)\cos 3\theta-8i\sigma (\Theta\pm \sigma)\sin\theta] \big )/\Theta,\\
			D_{\sigma,\pm 1}=&\gamma \big (4 A_\times (\Theta\mp \sigma)[4i\sigma-4i\sigma\Theta\cos\theta+4i(\sigma\pm \Theta)\cos 2\theta\pm 4 \sigma \Theta\sin\theta+(\Theta\mp \sigma)\sin 2\theta]+A_+[16(2\mp\Theta\sigma-\Theta^2)\cos\theta\nonumber \\&-16\sigma \Theta(\sigma\mp\Theta)+8i\sigma\Theta(\Theta\mp \sigma)\sin2\theta\pm i(15\mp 26\sigma\Theta+11\Theta^2)\sin\theta+i(\pm 3-2\sigma\Theta\mp \Theta^2)\sin3\theta]\big )/(2\Theta),\\
			D_{\sigma,0}=&-2\gamma^2\big (8A_\times \sin\theta[\cos\theta+\Theta(-1-i\sigma\sin\theta)]+A_+(2\Theta+7\cos\theta-2\Theta\cos2\theta+\cos3\theta) \big )   /\Theta.
		\end{align}
	\end{widetext}
	In the paraxial approximation where $\gamma\ll 1$, we retain only the lowest order terms in $\gamma$, giving
	\begin{align}
		C_{\pm 3}=&\gamma^3 [4A_\times \cos\theta\pm i A_+(3+\cos 2\theta)]\sin\theta,\\
		C_{\pm 2}=&4 \gamma^2k_3[\mp 4i A_\times \cos\theta+A_+(3+\cos2\theta)]/k_g,\\
		C_{\pm 1}=&32  \gamma k_3(A_\times \pm iA_+\cos\theta)\sin\theta/k_g,\label{eq.C1}\\
		C_0=&-32 k_3A_+\sin^2\theta/k_g,\\
		D_{\sigma,\pm 3}=&2 i\gamma(1\mp \sigma)[ 4iA_\times \cos\theta\mp A_+(3+\cos2\theta)]\sin\theta,\\
		D_{\sigma,\pm 2}=&4(\sigma \mp 1)[4iA_\times \cos\theta\mp A_+(3+\cos2\theta)],\\
		D_{\sigma,\pm 1}=&\gamma (-1\pm \sigma)\big (16A_\times\sin^2(\theta/2)(\pm i+\sin\theta)\nonumber \\&-A_+[8\cos\theta+i(8i\pm 13 \sin\theta\mp 4\sin2\theta\nonumber \\&\pm \sin3\theta)]\big ),\\
		D_{\sigma,0}=&-2\gamma^2[4A_\times(-i\sigma +i\sigma\cos2\theta-2\sin\theta+\sin2 \theta)\nonumber \\&+A_+(2+7\cos\theta-2\cos2\theta+\cos3\theta)],
	\end{align}
	with the time dependent factor $f(t)=[1-e^{ik_gt(\cos\theta-1)}]/[64(\cos\theta-1)]$.
	
	In most circumstances, the GW frequency is much lower than the photon frequency, such that $k_g\ll k_3$. For a typical scenario involving photons with wavelength $\lambda=355$ nm and GWs at $100$ Hz, the ratio $k_g/k_3$ is approximately $10^{-13}$. Under this condition, the Dirac delta function in Eq. \eqref{eq.innerProBetween01} indicates that the variation in the wave vector $k_3$ is negligible. Consequently, we ignore the variation of the light wave vector and hereafter omit the explicit wave vector indices $k_3$ and $k_\perp$. We utilize the Dirac notation $\ket{\sigma,l}$ to denote a photon state with SAM $\sigma$ and OAM $l$. Within the paraxial approximation, the index $m$ relates to the OAM via $m=l+\sigma$. In our model, the interaction between the light field and the GW begins at $t=0$, meaning the state for $t>0$ is expressed as
	\begin{equation}
		\ket {\sigma,l}=b^\dagger_{\sigma,l+\sigma}\ket 0.
	\end{equation}
	After the interaction, we measure the photon state in Minkowski spacetime by projecting it onto the unperturbed modes $\mathbf A^{(0)}_{I}$. The probability amplitude for the photon to occupy a state $\ket{\sigma',l'+\sigma'}$ is then given by
	\begin{equation}
		\braket{\sigma',l'}{\sigma,l}=\bra 0 {\hat a}_{\sigma',l'+\sigma'}{\hat b}^\dagger_{\sigma,l+\sigma}\ket 0=\alpha_{\sigma,l+\sigma,\sigma',l'+\sigma'}.
	\end{equation}
	Accordingly, the quantum state $\ket {\sigma,l}$ evolves within the GW as
	\begin{align}
		&\ket {\sigma,l}\nonumber \\
		\rightarrow &\ket {\sigma,l}+f(t)\sum_{j=-3}^3\left [C_{j}\ket {\sigma,l+j}+D_{\sigma,j}\ket {-\sigma,l+j+2\sigma} \right ].\label{eq.StateInGWs}
	\end{align}
	The state on the right hand side of Eq. \eqref{eq.StateInGWs} is unnormalized. However, the coefficients $C_{j}$ and $D_{\sigma,j}$ are proportional to the extremely small GW amplitudes $A_\lambda$. As a result, the normalization factor remains approximately unity and its variation is neglected in this paper.

	\section{Photon transition induced by GWs}
	
	In the preceding section, we demonstrated that photons undergo transitions to different quantum states when interacting with GWs. We identify the initial photons as source photons and the transitioned photons as signal photons. We define $P_{\Delta \sigma,j}$ as the transition probability from the initial state $\ket {\sigma,l}$ to the state $\ket {\sigma+\Delta \sigma,l+j-\Delta \sigma}$, where $\Delta \sigma \in \{-2,0,2\}$ represents the change in SAM and $j \in \{-3, \dots, 3\}$ denotes the change in total AM. Accordingly, the transition probability from the state $\ket {\sigma,l}$ to the state $\ket {\sigma,l+j}$ is expressed as
	\begin{equation}
		P_{0,j}=\left | f(t)C_{j} \right |^2,
	\end{equation}
	while the probability for a transition to the state $\ket {-\sigma,l+j+2\sigma}$ is given by
	\begin{equation}
		P_{-2\sigma,j}=\left |  f(t)D_{\sigma,j}   \right |^2.
	\end{equation}
	If a light field is emitted by a laser with power $Q$ and photon frequency $\omega$, the resulting transition photon rate, defined as the number of signal photons produced per unit time, is written as $\mathcal{N}=P_{\Delta \sigma,j}Q/\omega$ for any specific transition probability $P_{\Delta \sigma,j}$.
	
\begin{figure} [tbhp]
		\centering
		\includegraphics[width=1\linewidth]{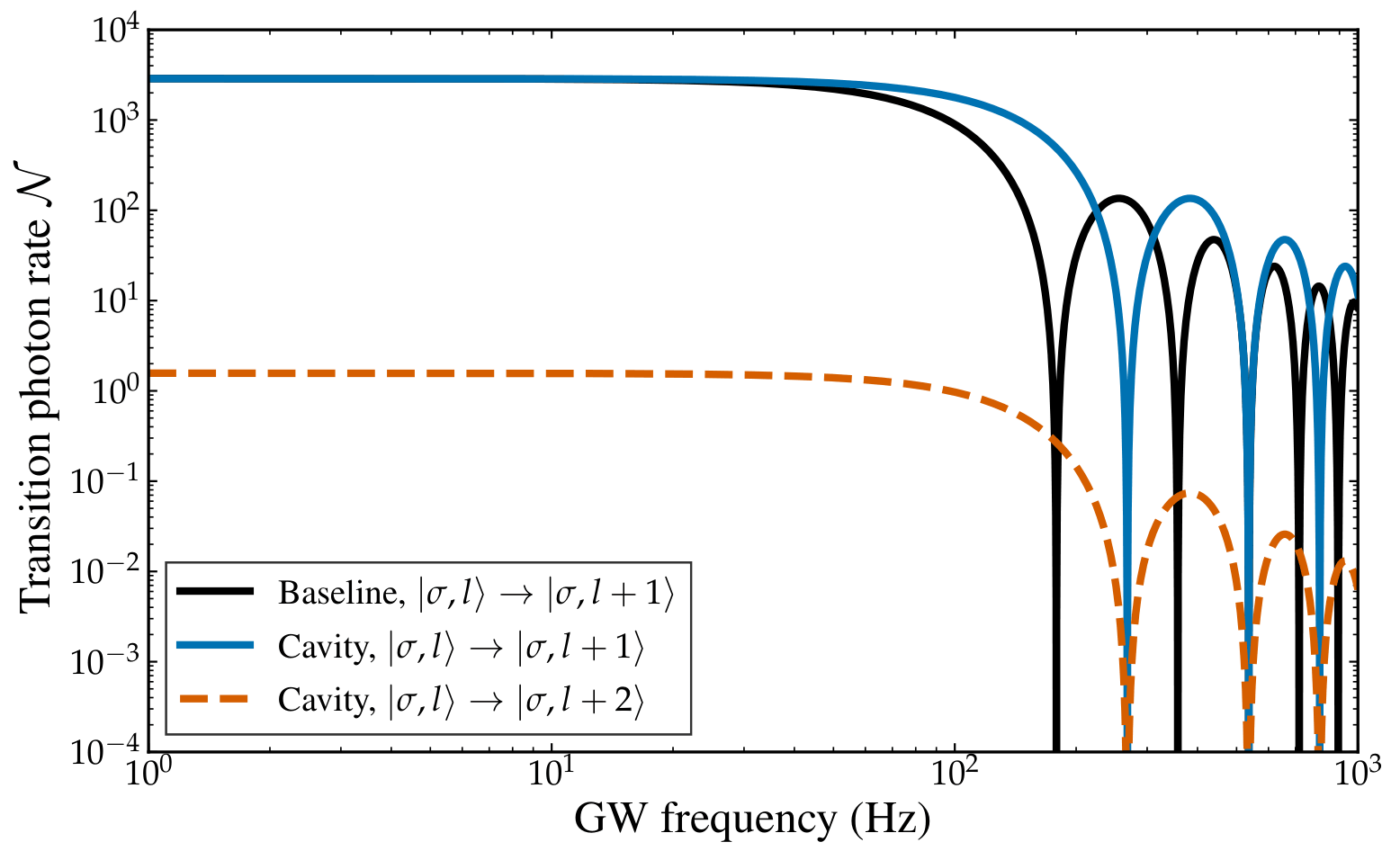}
		\caption{Comparison of transition photon rates $\mathcal N$ across the GW spectrum. The baseline case (black solid line) denotes the $\ket {\sigma,l} \rightarrow \ket {\sigma,l+ 1}$ transition with a $1120$ km propagation distance. The rates for the cavity configuration (4 km arm length, 140 round trips) are illustrated for the $\ket {\sigma,l} \rightarrow \ket {\sigma,l+ 1}$ transition (blue solid line) and the $\ket {\sigma,l} \rightarrow \ket {\sigma,l+ 2}$ transition (orange dashed line).}
		\label{fig.NvsFrequency}
	\end{figure}

	We investigate the transition probability for photons under two different configurations. The first scenario, defined as the baseline, involves photons detected after propagating over a large distance without any reflection. In Fig. \ref{fig.NvsFrequency}, the black solid line illustrates the transition photon rate, which is the number of signal photons produced per unit time, for the transition from $\ket {\sigma,l}$ to $\ket {\sigma,l+ 1}$ across a GW frequency range from 1 Hz to $10^3$ Hz. The simulation parameters include a photon wavelength of $\lambda=355$ nm, a radial wavevector of $k_\perp=1\times 10^6 ~{\rm m}^{-1}$, and GW strains of $A_+=A_\times=1\times10^{-21}$. We assume a laser power of $Q=5500$ W enhanced by a power recycling cavity, a GW incidence angle of $\theta=2\pi/3$, and an interaction distance of $L=1120$ km (the effective distance traveled by the laser in LIGO). In the low frequency regime, the system yields approximately $2880$ signal photons per unit time. Meanwhile, a dip occurs near 178 Hz, where the GW phase evolves by exactly $2\pi$ over the propagation length. Suppose at a point $d$ in the first half of the distance, a signal photon is induced with a phase $\phi(d)$ such that
	\begin{equation}
		e^{i\phi(d)}\ket {\sigma,l+1},~~0<d<L/2.
	\end{equation}
	At the corresponding point $d'=d+L/2$, the GW phase has increased by $\pi$, and the signal photon induced at this point is expressed as
	\begin{equation}
		e^{i[\phi(d)+\pi]}\ket {\sigma,l+1},~~d'=d+L/2.
	\end{equation}
	These two states interfere destructively, reducing the total number of signal photons. As the GW frequency increases beyond 178 Hz, new signal photons accumulate without this destructive interference, causing the transition photon rate $\mathcal N$ to rise.
	
\begin{figure} [tbhp]
		\centering
		\includegraphics[width=1\linewidth]{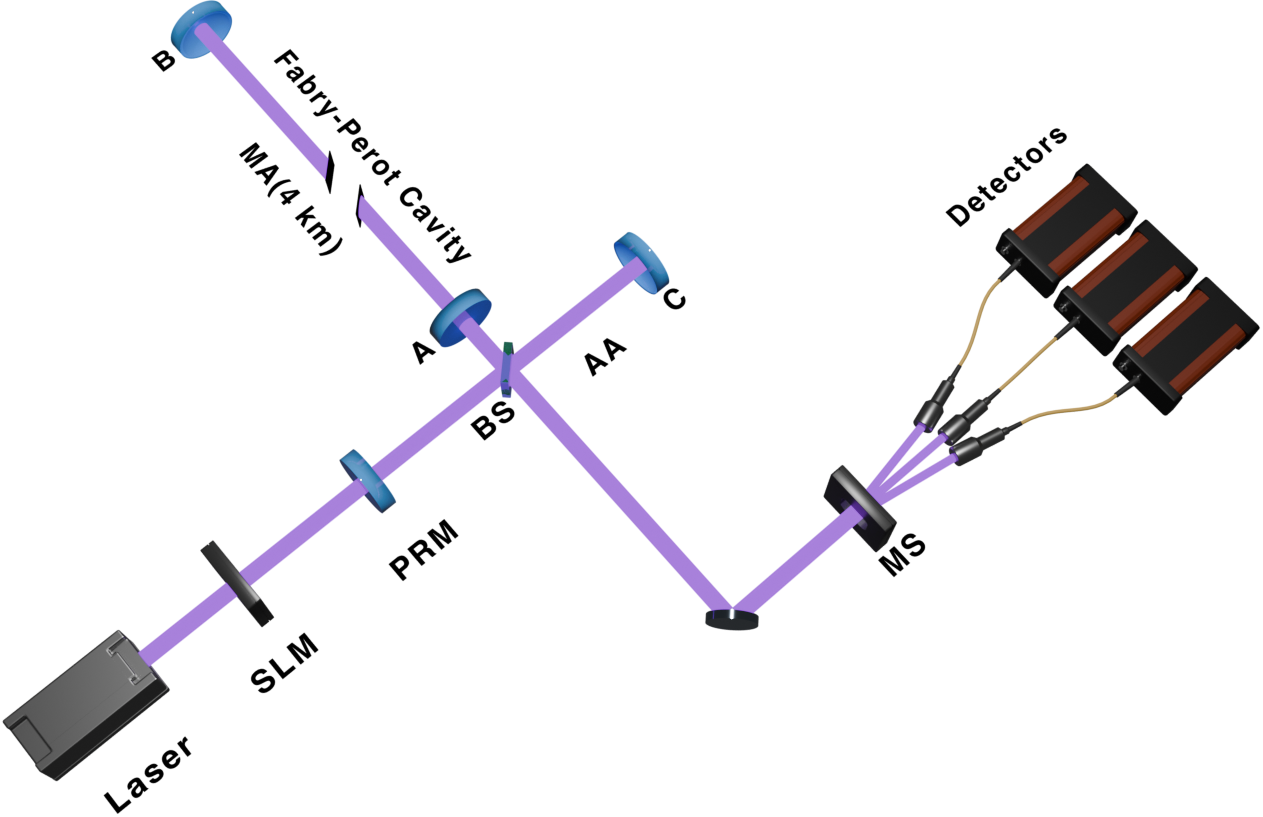}
		\caption{Schematic of a GW detector using quantum transitions in vortex beams. Component abbreviations include the spatial light modulator (SLM), power recycling mirror (PRM), beamsplitter (BS), main arm (MA), auxiliary arm (AA), and mode sorter (MS).}
		\label{fig.LightPath}
	\end{figure}

	Our study relies on the concept of a divergence-free Bessel beam. However, because an ideal Bessel beam requires infinite energy and is physically unrealizable, practical alternatives, including Bessel-Gaussian or Laguerre-Gaussian modes, are typically employed. Although these modes possess well-defined OAM, they experience significant divergence when propagating beyond the Rayleigh distance. This divergence makes it challenging to maintain a tightly localized light field over the $L=1120$ km baseline. To overcome this limitation, we fold the propagation distance into a cavity structure, as illustrated in Fig. \ref{fig.LightPath}. This configuration incorporates a spatial light modulator (SLM) for state preparation and a power recycling mirror (PRM) to amplify the laser source. The light is evenly divided by a beamsplitter (BS), with half of the source photons directed into a 4 km Fabry-Perot cavity. This cavity serves as the main arm (MA), where the light completes $N=140$ round trips. The laser and PRM are configured such that the power injected into the MA reaches 5500 W. The remaining half of the photons are routed into the auxiliary arm (AA). By precisely tuning the length of the AA, we induce complete destructive interference at the mode sorter (MS) port upon recombination at the BS. This effectively filters out the background source photons, allowing only the signal photons generated by GWs to be characterized by the MS and detectors. Also, because the photons in the AA are not required to interact with the GWs, the length of the AA can be kept minimal, requiring only a single reflection at mirror C.

	Before evaluating the transition probabilities, we must extend the transition amplitudes to the cavity configuration. Since the variation of the wave vector is ignored, all photons share an identical phase factor $k_\mu x^\mu + n_r \pi$, independent of whether a transition occurs. Here, the first term accounts for the propagation phase, the second term represents the phase shift due to reflection, and $n_r$ denotes the number of reflections. We therefore neglect this global phase and focus exclusively on the phase induced by the GWs as defined in Eq. \eqref{eq.StateInGWs}. Each of the $N$ round trips consists of a forward path from mirror A to mirror B and a backward path from mirror B to mirror A as shown in Fig. \ref{fig.LightPath}. For the $n$th forward trip ($A\rightarrow B$), we define the origin at point A with the $z$ axis directed toward B. At this starting point, the GW possesses a phase $\phi_0=-2(n-1)k_g L$. The resulting transition amplitude is
	\begin{equation}
		\alpha^{\rm (A\rightarrow B)}_{\sigma,l+\sigma,\sigma',\sigma'+l'}=e^{-2i(n-1)k_gL}\alpha_{\sigma,l+\sigma,\sigma',\sigma'+l'}(A_+,A_\times,\theta),
	\end{equation}
	where we explicitly write the dependence on the GW parameters. For the backward path ($B\rightarrow A$), we introduce a new coordinate system $(x',y',z')$ with the origin at point B and the $z$ axis reversed such that $x'=x$, $y'=-y$, and $z'=-z+L$. In this frame, the polarization amplitudes transform to $A'_+=A_+$ and $A'_\times=A_\times$ while the propagation angle becomes $\theta'=\pi-\theta$. At the beginning of the $n$th backward trip, the GW picks up a phase $\phi_0=-(2n-1)k_gL+k_g L \cos\theta$ at point B. Additionally, the SAM and OAM are reversed upon reflection. The backward transition amplitude is thus given by
	\begin{align}
		\alpha^{\rm (B\rightarrow A)}_{\sigma,l+\sigma,\sigma',\sigma'+l'}=&e^{-i(2n-1)k_gL+ik_g L \cos\theta}\nonumber \\ &\times \alpha_{-\sigma,-l-\sigma,-\sigma',-\sigma'-l'}(A_+,A_\times,\pi-\theta).
	\end{align}
	The total transition amplitude after $N$ round trips is the sum
	\begin{equation}
		\alpha^{\rm (cavity)}_{\sigma,l+\sigma,\sigma',\sigma'+l'}=\sum_{n=1}^N \left [\alpha^{\rm (A\rightarrow B)}_{\sigma,l+\sigma,\sigma',\sigma'+l'}+\alpha^{\rm (B\rightarrow A)}_{\sigma,l+\sigma,\sigma',\sigma'+l'} \right ],
	\end{equation}
	yielding the probability $P_{\Delta \sigma,j}=\left | \alpha^{\rm (cavity)}_{\sigma,l+\sigma,\sigma+\Delta \sigma,l+\sigma+j} \right |^2$. In Fig. \ref{fig.NvsFrequency}, the blue solid line and orange dashed line represent the transition photon rates for $\ket {\sigma,l}\rightarrow \ket {\sigma,l+ 1}$ and $\ket {\sigma,l}\rightarrow \ket {\sigma,l+ 2}$, respectively. While the cavity configuration yields a smaller transition photon rate than the baseline for the $\ket {\sigma,l}\rightarrow \ket {\sigma,l+ 1}$ transition, its first dip at 267 Hz is higher. Consequently, the cavity configuration enables signal detection across a broader frequency spectrum. The $\ket {\sigma,l}\rightarrow \ket {\sigma,l+ 2}$ transition produces much fewer photons because the coefficient $C_{\pm 2}$ is suppressed by the paraxial parameter $\gamma$ relative to $C_{\pm 1}$. Furthermore, the $\ket {\sigma,l}\rightarrow \ket {\sigma,l\pm 3}$ transition is suppressed by both $\gamma^2$ and the small factor $k_g/k_3$, rendering only the $\ket {\sigma,l}\rightarrow \ket {\sigma,l\pm 1}$ transition practical for GW detection.
	
\begin{figure} [tbhp]
		\centering
		\includegraphics[width=1\linewidth]{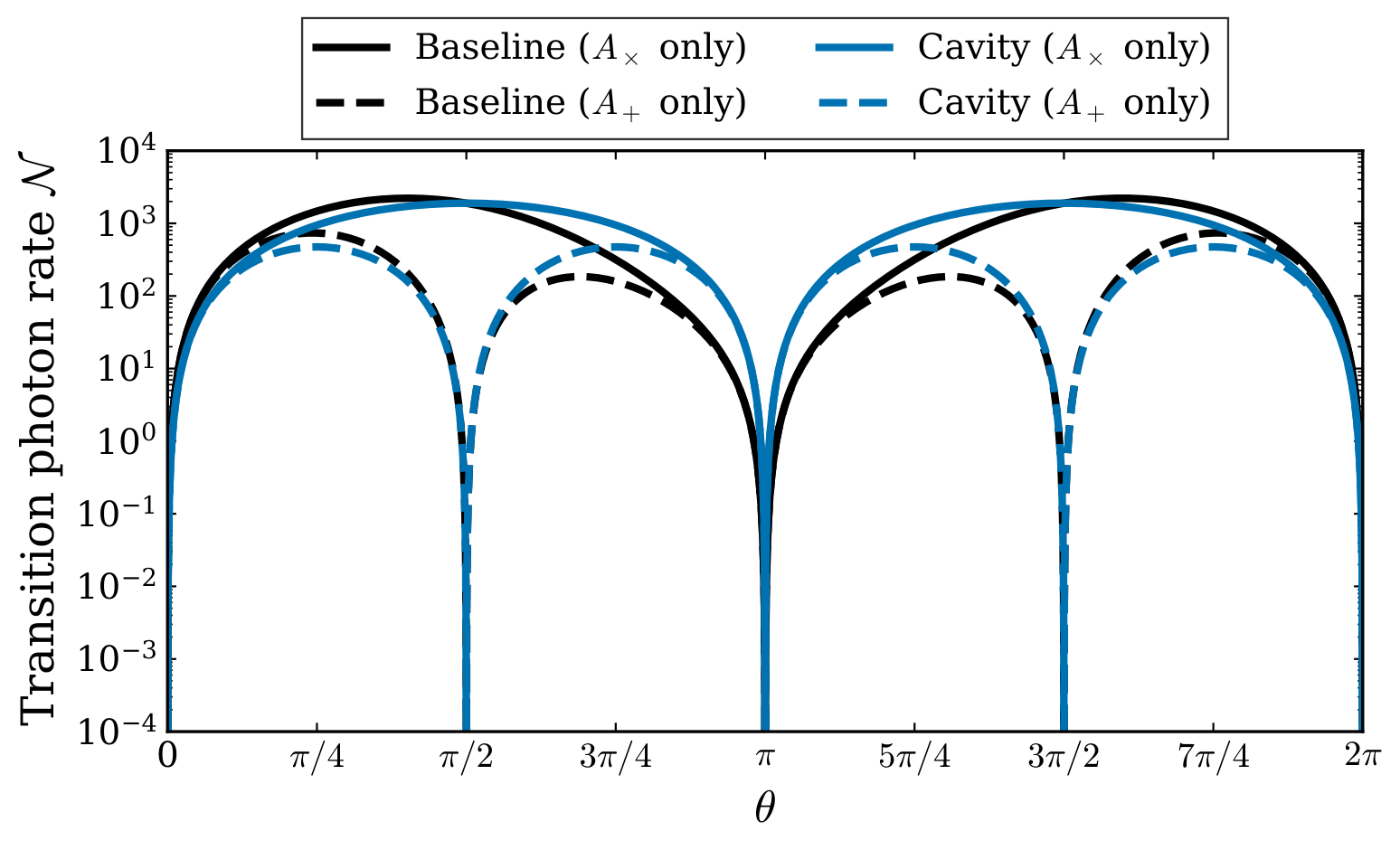}
		\caption{Transition photon rate $\mathcal N$ across the angular spectrum of GW propagation. With the frequency set at 100 Hz, the plot compares the baseline (black) and cavity (blue) responses. The $A_\times$ polarization effects are shown with solid lines, and the $A_+$ polarization effects are depicted with dashed lines.}
		\label{fig.NvsTheta}
	\end{figure}

	The transition amplitudes depend explicitly on the GW polarization amplitudes $A_+$ and $A_\times$ as well as the propagation direction $\theta$. In Fig. \ref{fig.NvsTheta}, we plot the transition photon rate for $\ket {\sigma,l}\rightarrow \ket {\sigma,l+ 1}$ as a function of these parameters with the GW frequency fixed at 100 Hz. For the baseline configuration, represented by the black lines, the polarization amplitudes $A_+$ and $A_\times$ exert impacts of similar magnitude, except at $\theta=\pi/2$ and $3\pi/2$ where the effect of $A_+$ vanishes. This behavior reflects the dependence of the coefficient $C_\pm 1$ in Eq. \eqref{eq.C1} on $A_+\cos\theta$. In the cavity configuration, indicated by the blue lines in Fig. \ref{fig.NvsTheta}, the contribution from $A_+$ at $\theta=\pi/2$ and $3\pi/2$ is significantly suppressed, as well. When a photon transitions to the state $\ket {\sigma,l+ 1}$ along the forward path ($A\rightarrow B$), the associated probability amplitude scales as $C_1 \propto (A_\times+iA_+\cos\theta)\sin\theta$. Conversely, along the backward path ($B\rightarrow A$), the SAM and OAM of the source photons are reversed to $-\sigma$ and $-l$, respectively, and the signal photon has the OAM of $-l-1$. This transition is governed by the coefficient $C_{-1}$, where $C_{-1} \propto (A'_\times-iA'_+\cos\theta')\sin\theta'=(A_\times+iA_+\cos\theta)\sin\theta$. Therefore, the signal photons generated along the backward path interfere constructively with those from the forward path, effectively enhancing the GW effects on photons.
	
\begin{figure} [tbhp]
		\centering
		\includegraphics[width=1\linewidth]{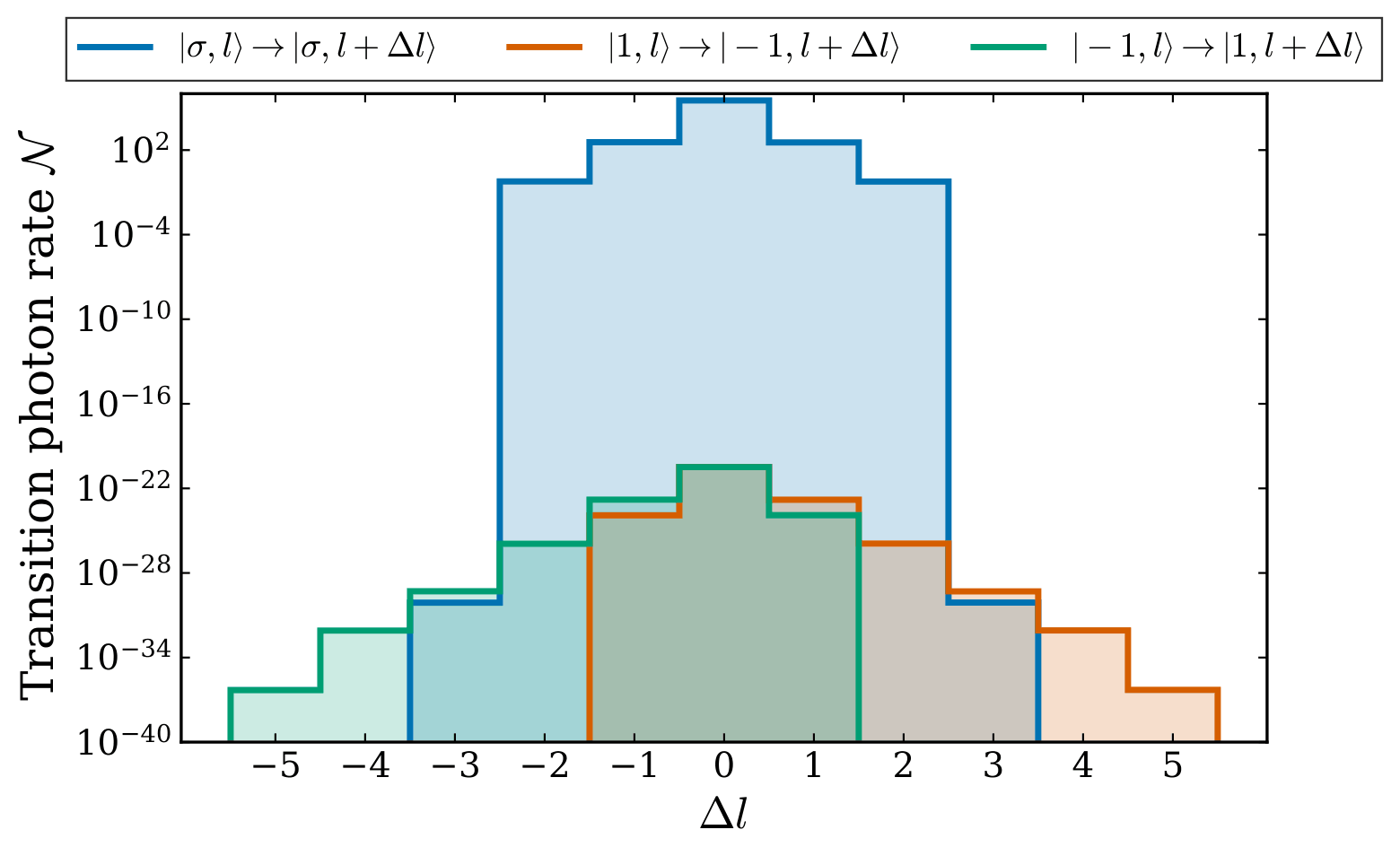}
		\caption{Transition photon rates $\mathcal N$ across fourteen different quantum transition pathways. The blue area illustrates transitions with conserved SAM. Chirality flip transitions are indicated by the orange area for initial state $\sigma=1$ and the green area for initial state $\sigma=-1$.}
		\label{fig.NvsCD}
	\end{figure}

	Although our study primarily focuses on the $\ket {\sigma,l} \rightarrow \ket {\sigma,l\pm 1}$ transition, the coefficients $C_j$ and $D_{\sigma,j}$ define fourteen different transition channels, the majority of which exhibit negligible probabilities. Figure \ref{fig.NvsCD} displays the transition photon rate $\mathcal N$ for the complete set of transition channels within the cavity configuration. The GW parameters are fixed at strains $A_+=A_\times=1\times10^{-21}$, an incident angle of $\theta=2\pi/3$, and a frequency of $f=100$ Hz. The blue region denotes quantum transitions characterized by conserved SAM. The rate $\mathcal N$ is dominated by the $\Delta l=0$ and $\pm 1$ transitions, with substantial suppression observed as $\left |\Delta l \right |$ grows. A key distinction from scalar OAM models \cite{q38s-k2tq} is the emergence of the $\ket {\sigma,l}\rightarrow \ket {\sigma,l\pm 3}$ channels, which are unique to the vector EM field. Because these channels are driven by first-order derivatives of the GWs, they are heavily suppressed.
	
	The observed suppression of spin-flip transitions relative to OAM transitions arises from the differences between local and topological properties. Unlike SAM, which is defined locally, OAM is a topological property determined by the azimuthal phase winding $l=\oint_C \nabla \phi(\mathbf r)\cdot d\mathbf r/(2\pi)$ around the propagation axis \cite{allen1992orbital,Gbur:08}. This topological structure renders the OAM sensitive to gravitational perturbations that accumulate over a finite spatial region. In contrast, the SAM is a local property characterized by the polarization state at a specific coordinate $x^i$ \cite{PhysRevLett.88.053601}. A polarization detector at $x^i$ can determine the SAM without reference to the beam center or any distant points. Because a single massless particle cannot locally perceive a GW due to the existence of local transverse-traceless coordinates where it remains stationary \cite{flanagan2005basics}, observing physical GW effects requires nonlocal gradients. In our framework, this nonlocality enters through terms involving the first derivatives of the metric perturbation $h^{\alpha\beta}_{~~,\mu}$. These derivative terms introduce the GW vector $k_g$ into the EOMs, effectively suppressing the $\ket {\sigma,l}\rightarrow \ket {-\sigma,l'}$ transition probability by a factor of $(k_g/k)^2$. Thus, the photon SAM remains a robust intrinsic property during GW interaction, while the OAM, governed by the spatial phase distribution, acts as the primary channel for gravitational sensing. This behavior is shown in the data presented in Fig. \ref{fig.NvsCD}.
	
	The interaction between the light field and GWs is characterized by a total AM shift bounded within $\pm 3$ under the $k_g, k_\perp \ll k_3$ approximation. Although additional channels might emerge if these conditions are broken, their contributions are negligible. We observe that transitions with conserved SAM result in an OAM spectrum that is symmetric about $\Delta l=0$, a consequence of the nonrotating nature of the $A_+$ and $A_\times$ linear polarizations. This symmetry is broken if one adopts the circular polarization basis where $A_+=(A_R+A_L)/\sqrt 2$ and $A_\times=i(-A_R+A_L)/\sqrt 2$ \cite{misner1973gravitation}. Under this representation, the amplitudes $A_R$ and $A_L$ for right and left circular GWs introduce a preferred rotation, leading to asymmetric transition probabilities \cite{q38s-k2tq}. Furthermore, if the SAM undergoes a reversal from $-1$ to $1$, the OAM shift $\Delta l$ is restricted to the range from $-5$ to $1$. This shift produces an asymmetric OAM spectrum that reflects the underlying coupling between the SAM and OAM during the gravitational interaction.

	\section{Conclusion}
	
	In this study, we employed perturbation theory to analyze the coupling between vector Bessel beams and GWs. By calculating transition photon rates across diverse parameter spaces, we demonstrated that the OAM of light is sensitive to gravitational perturbations, compared to the SAM, which remains largely invariant due to its local definition at specific spatial coordinates. Because the OAM is a topological property determined by the phase winding around the propagation axis, it naturally accumulates gravitational effects over a finite spatial volume. Conversely, the SAM only interacts with GWs through the spatial derivatives of the metric, which introduces a suppression factor proportional to the ratio $k_g/k_3$. Our findings confirm that while the SAM acts as a robust intrinsic quantum number, the OAM serves as a highly responsive probe for GW detection. This interaction highlights the potential of vortex light as a high precision tool in GW astronomy.
	
	Despite the theoretical existence of fourteen transition channels, the $\ket {\sigma,l}\rightarrow \ket {\sigma,l\pm 1}$ pathways provide the most viable mechanism for experimental GW detecting. This detection strategy can use the intensity profiles of OAM states, where nonzero OAM beams possess a central dark intensity null due to their helical phase singularities \cite{shen2019optical}. In contrast, the $l=0$ state is the only mode that exhibits a bright intensity maximum at the beam center \cite{shen2019optical}. By preparing the initial photon state with $l=1$, the GW interaction populates the $l=0$ mode for a subset of the photons. This transition effectively fills in the central dark core, producing a detectable bright signal at the axis of propagation. This approach allows for high sensitivity measurements by searching for a bright signal against a nominally dark background.
	
	
	Although a noise budget analysis is beyond the scope of the current work, we identify the key quantum and thermal factors that will govern the sensitivity of the detector. Unlike traditional interferometers, our method is inherently insensitive to fluctuations in the arm length, allowing it to bypass seismic noise and other displacement based background noise \cite{pitkin2011gravitational}. The dominant limitation is expected to be shot noise from the discrete nature of photon detection. In current GW observatories, reducing shot noise by increasing circulating power leads to a proportional increase in radiation pressure noise, which perturbs the test mass positions \cite{pitkin2011gravitational}. Given that our signal is encoded in the quantum transitions of the light field rather than the displacement of arm lengths, radiation pressure is not expected to be the limiting noise source. Since the current work focuses on the interaction theory between vector light fields and GWs, further studies involving a detailed noise budget analysis are required before this technique can be experimentally realized.

	\section*{Acknowledgments}
	This work is supported by National Natural Science Foundation of China (125B2103).
	
	\section*{Data availability}
	The data that support the findings of this article are not publicly available. The data are available from the authors upon reasonable request.
	
	\appendix
	
	\section{Useful equations and identities}
	We provide a summary of the equations and identities concerning Dirac delta functions and Bessel functions. Throughout this paper, the superscript $(n)$ applied to a function $f(x)$ denotes its $n$th derivative, defined as $f^{(n)}(x)=d^nf(x)/dx^n$. These relations are useful to the perturbative analysis of the vector Bessel beam interaction with GWs.
	
	(a) We provide several identities for the Dirac delta functions used throughout the preceding derivations.
	\begin{equation}
		\int_{-\infty}^\infty dz (\alpha z+\beta)e^{ikz}=-2\pi i \alpha \delta^{(1)}(k)+2\pi \beta \delta(k),
	\end{equation}
	\begin{equation}
		f(k_\perp,k'_\perp)\delta^{(1)}(k_\perp,k'_\perp)=\frac{df(k_\perp,k'_\perp)}{dk'_\perp}\delta (k_\perp,k'_\perp),
	\end{equation}
	\begin{equation}
		\delta^{(1)}(-x)=-\delta^{(1)}(x),
	\end{equation}
	\begin{equation}
		k_\perp \delta^{(1)}(k_\perp,k'_\perp)+k'_\perp\delta^{(1)} (k_\perp,k'_\perp)=-\delta  (k_\perp,k'_\perp).
	\end{equation}
	
	(b) Numerous integrals involving Bessel functions appear throughout this paper and the following identities are provided to facilitate their evaluation (some can be found in Refs. \cite{PhysRevA.71.033411,lozier2003nist,olver2010nist}).
	\begin{align}
		&(k_\perp \rho)^2J^{(2)}_m(k_\perp \rho)+k_\perp \rho J^{(1)}_m(k_\perp \rho)\nonumber \\&+[(k_\perp \rho)^2-m^2]J_m(k_\perp \rho)	=0,
	\end{align}
	\begin{align}
		&(k_\perp \rho)^2J^{(3)}_m(k_\perp \rho)+3k_\perp \rho J^{(2)}_m(k_\perp \rho)+[k^2_\perp \rho^2-m^2+1]\nonumber \\ &\times J_m^{(1)}(k_\perp \rho)+2 k_\perp \rho J_m(k_\perp \rho)	=0,
	\end{align}
	\begin{equation}
		J_{m+1}(k_\perp \rho)=\frac {m}{k_\perp \rho}J_m(k_\perp \rho)-J^{(1)}_m(k_\perp \rho),
	\end{equation}
	\begin{equation}
		J_{m-1}(k_\perp \rho)=\frac {m}{k_\perp \rho}J_m(k_\perp \rho)+J^{(1)}_m(k_\perp \rho),
	\end{equation}
	\begin{equation}
		\int_0^\infty d\rho \rho J_m(k_\perp \rho)J_m(k'_\perp \rho)=\frac 1 {k_\perp}\delta(k_\perp-k'_\perp),
	\end{equation}
	\begin{align}
		&\int_0^\infty d\rho \rho^2 J_m(k_\perp \rho)J_m^{(1)}(k'_\perp \rho)\nonumber \\
		=&-\frac{\delta^{(1)}(k_\perp-k'_\perp)}{k'_\perp}-\frac {\delta(k_\perp-k'_\perp)} {k'^2_\perp},
	\end{align}
	\begin{align}
		&\int_0^\infty d\rho\left [\frac {m^2}{k_\perp k'_\perp \rho }J_m(k_\perp \rho)J_m(k'_\perp \rho)+\rho   J_m^{(1)}(k_\perp \rho)J_m^{(1)}(k'_\perp \rho) \right ]  \nonumber \\
		=&\frac {\delta(k_\perp -k'_\perp)}{k_\perp},
	\end{align}
	\begin{align}
		&\int_0^\infty d\rho\left [k_\perp  J_m^{(1)}(k_\perp \rho)J_m(k'_\perp \rho)+k'_\perp   J_m(k_\perp \rho)J_m^{(1)}(k'_\perp \rho) \right ]  \nonumber \\
		=&-J^2_m(0),
	\end{align}
	\begin{equation}
		m J^2_m(0)=0.
	\end{equation}
	
	(c) The evaluation of the following integral,
	\begin{align}
		F=&\int_0^\infty d\rho \rho[(a+b)J_{m+1}(k'_\perp \rho)J_{m-1}(k_\perp \rho)\nonumber \\&+(a-b) J_{m+1}(k_\perp \rho)J_{m-1}(k'_\perp \rho)],
	\end{align}
	for $m\le 0$ requires careful consideration. We begin with the identity
	\begin{equation}
		I=\int_0^\infty d\rho J_m(\alpha\rho)J_{m-1}(\beta\rho)=\begin{cases}
			\beta^{m-1}\alpha^{-m},&\beta<\alpha,\\
			(2\beta)^{-1},&\beta=\alpha,\\
			0,&\beta>\alpha.
		\end{cases}
	\end{equation}
	By applying recurrence relations, this integral is expressed as
	\begin{align}
		I=&\int_0^\infty \frac{d\rho}{2m} [J_{m-1}(\alpha\rho)+J_{m+1}(\alpha\rho)]J_{m-1}(\beta\rho)\nonumber \\
		=&\frac {\alpha }{2m}\left [\frac {\delta(\alpha-\beta)}{\beta}+\int_0^\beta d\rho\rho J_{m+1}(\alpha\rho)J_{m-1}(\beta\rho)\right ].
	\end{align}
	Consequently, we obtain the relation
	\begin{align}
		&\int_0^\infty d\rho \rho J_{m+1}(\alpha \rho)J_{m-1}(\beta \rho)\nonumber \\
		=&\begin{cases}
			2m \beta^{m-1}\alpha^{-m-1},&\beta<\alpha,\\
			m(\alpha\beta)^{-1}-\delta(\alpha-\beta)\beta^{-1},&\beta=\alpha,\\
			0,&\beta>\alpha.
		\end{cases}
	\end{align}
	In the case where $k_\perp>k'_\perp$, the integral $F$ simplifies to
	\begin{equation}
		F=(a-b)\frac {2m}{k_\perp k'_\perp}\left ( \frac{k'_\perp}{k_\perp}\right )^m.
	\end{equation}
	If $m=0$, then $F=0$. Otherwise, since $k'_\perp/k_\perp<1$ and the final result involves a subsequent integration with respect to $k'_\perp$ from $0$ to $k_\perp$, this contribution is strongly suppressed and may be safely ignored. Thus we take $F\approx 0$. A similar derivation yields $F\approx 0$ for $k_\perp<k'_\perp$. For the case where $k_\perp=k'_\perp$, the expression becomes
	\begin{equation}
		F=2a[mk^{-2}_\perp-k^{-1}_\perp\delta(k_\perp-k'_\perp)]\approxeq \frac {-2a}{k_\perp}\delta(k_\perp-k'_\perp),
	\end{equation}
	Here, the first term is neglected because the Dirac delta function dominates at the point $k\perp=k'\perp$. Therefore, the final result of the integral is given by
	\begin{equation}
		F={-2a}{k_\perp}\delta(k_\perp-k'_\perp).
	\end{equation}

\end{document}